\documentclass{elsart}
\usepackage{graphicx}
\usepackage{epsfig}
\usepackage{amssymb}
\newcommand{\be}{\begin{equation}}
\newcommand{\ee}{\end{equation}}

\begin{document}
\begin{frontmatter}
\title{The scale-free topology of market investments}
\author[1]{Diego Garlaschelli}
\author[2]{Stefano Battiston}
\author[3]{Maurizio Castri}
\author[4]{Vito D. P. Servedio}
\and
\author[4,5]{Guido Caldarelli}

\address[1]{INFM and Dipartimento di Fisica, Universit\`a di Siena, Via Roma 56, 53100 Siena ITALY}
\address[2]{Laboratoire de Physique Statistique ENS, 24 rue Lhomond, 75005 Paris FRANCE}
\address[3]{DYRAS02, Via L. Einaudi 10, 05017 Monteleone di Orvieto (TR) ITALY}
\address[4]{INFM Udr Roma ``La Sapienza'', P.le A. Moro 2, 00185 Roma ITALY}
\address[5]{Centro Studi e Ricerche Enrico Fermi, Compendio Viminale, 00184 Roma ITALY}
\begin{abstract}
We propose a network description of large market investments, where both stocks and shareholders are represented as vertices connected by weighted links corresponding to shareholdings. In this framework, the in-degree ($k_{in}$) and the sum of incoming link weights ($v$) of an investor correspond to the number of assets held (\emph{portfolio diversification}) and to the invested wealth (\emph{portfolio volume}) respectively. An empirical analysis of three different real markets reveals that the distributions of both $k_{in}$ and $v$ display power-law tails with exponents $\gamma$ and $\alpha$. Moreover, we find that $k_{in}$ scales as a power-law function of $v$ with an exponent $\beta$. Remarkably, despite the values of $\alpha$, $\beta$ and $\gamma$ differ across the three markets, they are always governed by the scaling relation $\beta=(1-\alpha)/(1-\gamma)$. We show that these empirical findings can be reproduced by a recent model relating the emergence of scale-free networks to an underlying Paretian distribution of `hidden' vertex properties.
\end{abstract}
\begin{keyword}
Complex Networks \sep Econophysics \sep Wealth Distribution \sep Pareto's Law
\PACS 89.75.-k \sep 89.65.Gh \sep 02.50.Ey \sep 87.23.Ge
\end{keyword}
\date{23 January 2004}
\end{frontmatter}

\section{Introduction}
A fundamental problem in economics is to characterize different systems by means of simple and universal features. The power-law form of the statistical distributions of many quantities, including individual wealth\cite{1,2,3,6}, firm size\cite{7} and financial market fluctuations\cite{10,11,13}, seems to be one of such `stylized facts'. As in many other complex systems, the emergence of this behaviour can be related to the interactions of a large number of agents\cite{16,17,18}. On the other hand, the recent advances in network theory\cite{19} allow to describe economic systems internally and to characterize them through novel quantities. Indeed, the topology of various economic networks, ranging from those formed by directors of corporate boards\cite{19,battiston} to those generated by the strongest asset correlations\cite{bonanno} is again characterized by power-law distributions, in close analogy with many other networks (including Internet, WWW and biological webs\cite{19}).

In the present paper we propose a network description of the financial system formed by the assets traded in a stock market and the corresponding shareholders. As we show below, we find that shareholding networks are again characterized by power-law distributions, which here describe the volume and diversification of portfolios. These quantities are the subject of fundamental financial issues such as portfolio optimization\cite{22}, and our empirical analysis reveals that they are related through nontrivial scaling relations. We finally show that the above results can be reproduced by a simple network model\cite{fitness} which assumes that the topological properties depend on some non-topological quantity (or \emph{fitness}) which is this case represents the wealth invested by the shareholders.

\section{Introducing the shareholding networks}
The data sets we analysed refer to the shareholders of all assets traded in the Italian stock market\cite{25} (MIB) in the year 2002, in the New York Stock Exchange\cite{26}(NYSE) in the year 2000 and in the National Association of Security Dealers Automated Quotations\cite{26} (NASDAQ) in the year 2000. The number $M$ of assets in the markets is 240, 2053 and 3063 respectively. The data necessarily report, for each asset, only a limited number of investors (generally holding a significant fraction of shares of it). While this biases the estimate of the number of investors of each asset (which can in principle be very large), it does not affect qualitatively the statistical properties of the number of assets in the portfolio of each reported investor.

\begin{figure}[]	
\centerline{\epsfxsize 5.2 truein \epsfbox{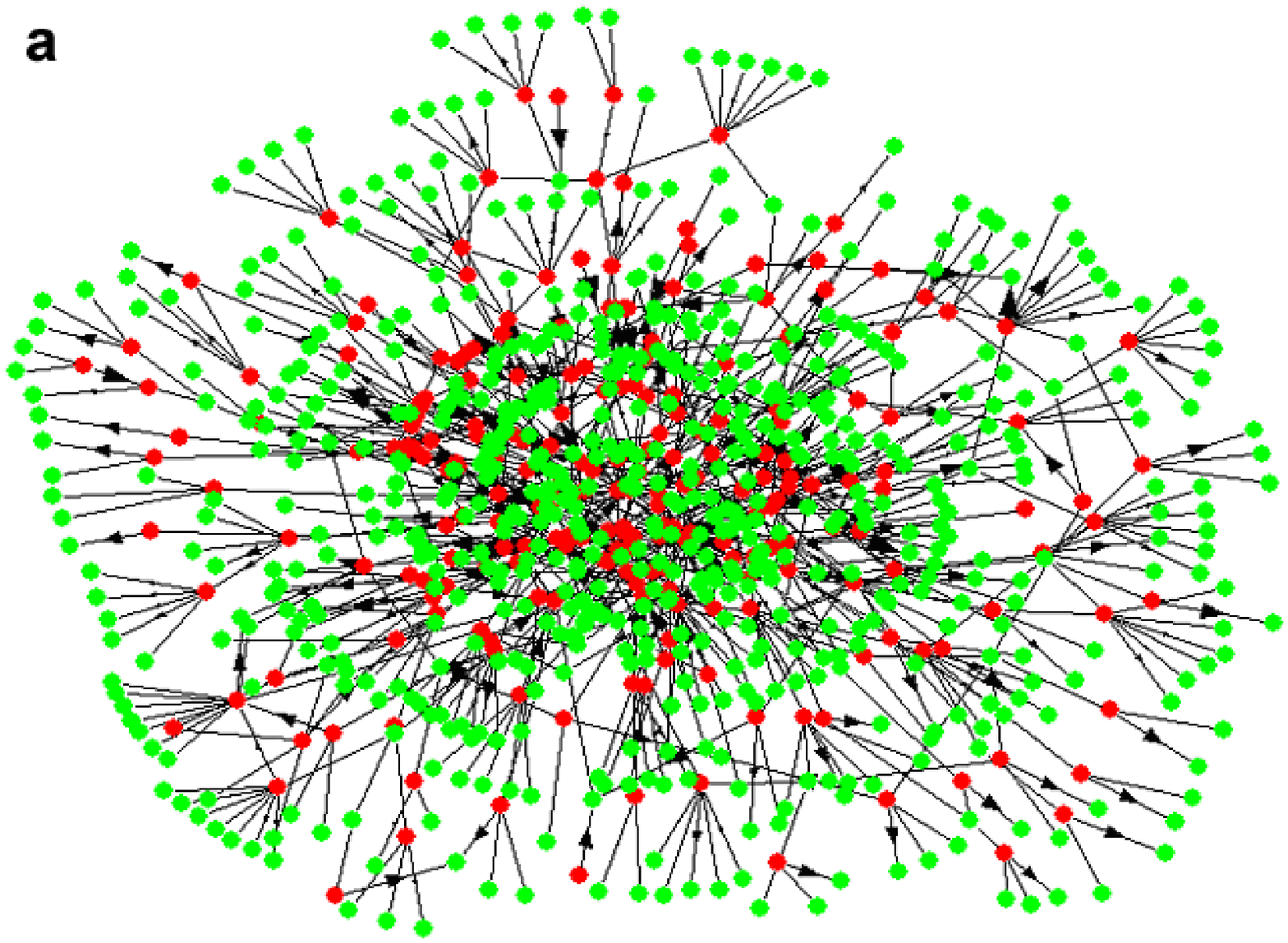}}
\centerline{\epsfxsize 5.2 truein \epsfbox{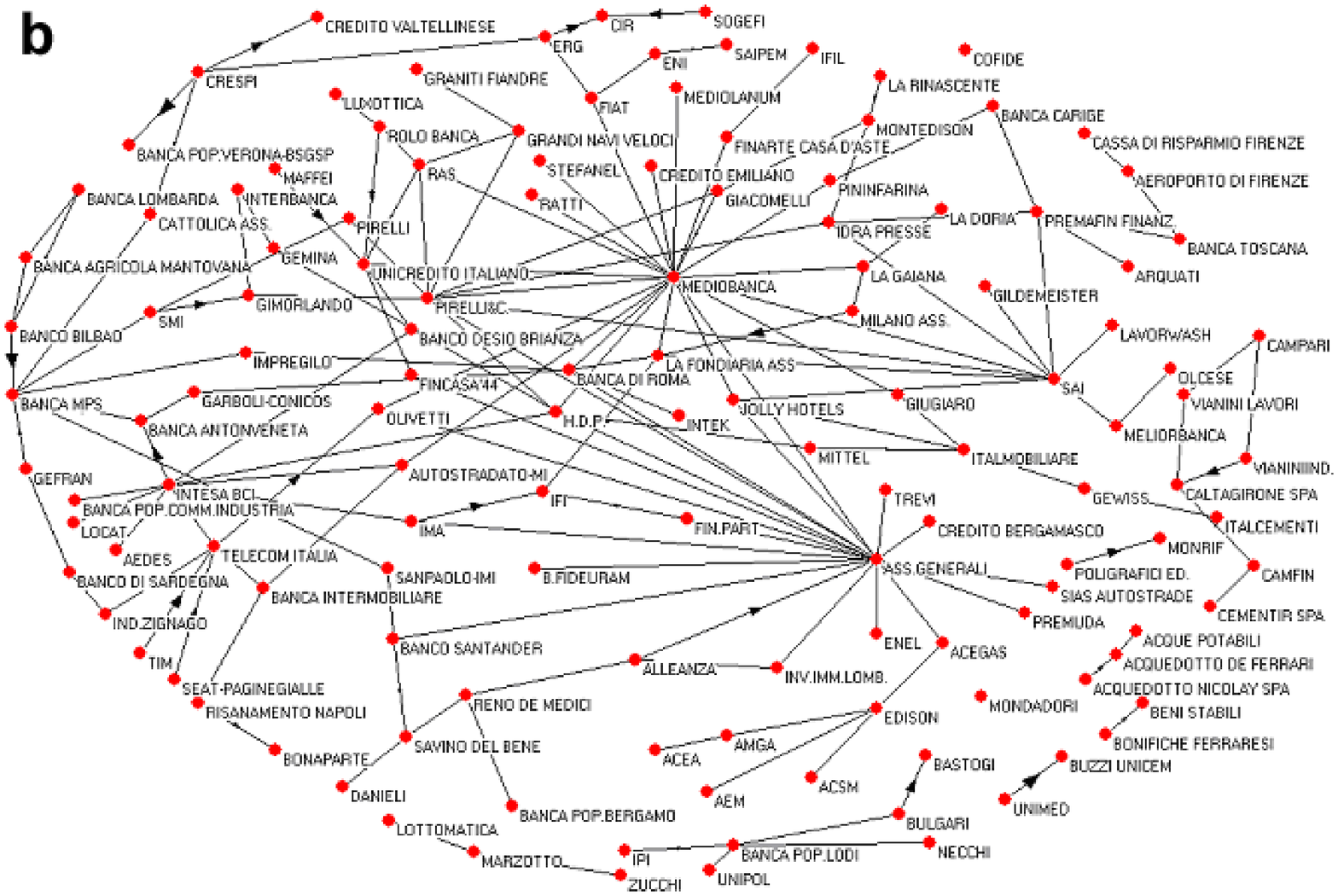}}
\caption[]{
\label{fig1}
\small Shareholding networks for the Italian market: a) the extended net (red vertices = stocks, green vertices = shareholders) and b) the restricted one (stocks labelled by the name of the corresponding company). Arrow size is proportional to the fraction of shares owned.}
\end{figure}

As well known, it happens frequently that some shareholders of a certain company are themselves companies whose shares are traded in the market, so that there is a significant fraction of listed companies which are also owners of other listed companies. This leads naturally to a network description of the whole system (see Fig.\ref{fig1}a), where the $N$ investors and the $M$ assets are both represented as vertices and a directed link is drawn from an asset to any of its shareholders (which can be persons or listed companies themselves, therefore the total number of vertices is less than $N+M$). In this topological description the in-degree $(k_{in})_i$ (number of incoming links) an the investor $i$ corresponds to the number of different assets in its portfolio (which we call the `portfolio diversification'). Vertices with zero in-degree are listed companies holding no shares of other stocks. The out-degree $k_{out}$ of a vertex is the number of shareholders of the corresponding asset, but as we discussed above this is a biased quantity and we cannot deal with its statistical description. We also note that a weight can be assigned to each link, defined as the fraction $s_{ij}$ of the shares outstanding of asset $j$ held by $i$ multiplied by the market capitalization $c_j$ of the asset $j$. The quantity $v_i=\sum_j s_{ij}c_j$ (hereafter the `portfolio volume') is therefore the total wealth in the portfolio of $i$. If we consider the subnet restricted to the owners which are listed companies themselves (hereafter the `restricted' net), we obtain the structure reported in Fig.\ref{fig1}b, providing a description of the interconnections among stocks. The whole networks will be denoted as the `extended' ones.

\begin{figure}[]	
\centerline{\epsfxsize 4.5 truein \epsfbox{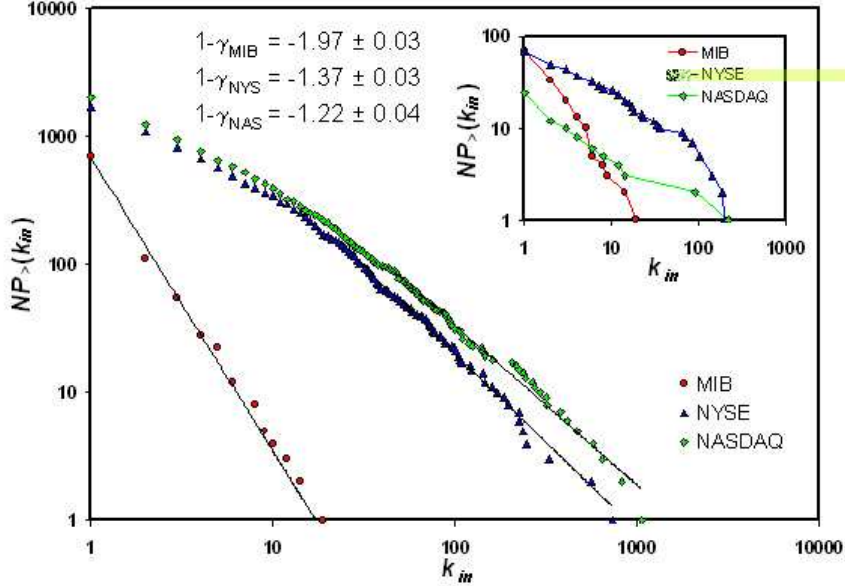}}
\caption[]{
\label{fig2}
\small Cumulative histograms of $k_{in}$ for the extended nets (mai panel, with power-law fit) and the restricted ones (inset).}
\end{figure}

In order to characterize the topology of these systems we consider the statistical distribution $P_>(k_{in})$ of the number of vertices with in-degree greater than or equal to $k_{in}$. This analysis has been performed on both the extended and the restricted nets. As reported in Fig.\ref{fig2}, the tail of the distribution computed on the extended nets can always be fitted by a power law of the form \be\label{pk} 
P_>(k_{in})\propto k_{in}^{1-\gamma}
\ee
This corresponds (for large values of $k_{in}$) to a probability density 
$P(k_{in})\propto k_{in}^{-\gamma}$ of finding a holder with a portfolio diversified in exactly $k_{in}$ different stocks. The values of the exponent $\gamma$ differ across markets: $\gamma_{NYS}=2.37$, $\gamma_{NAS}=2.22$, $\gamma_{MIB}=2.97$ (however note that in the Italian case the quite large exponent $\gamma_{MIB}$ and the small size of the net result in a small value $k_{in}^{max}=19$ of the largest degree). In the inset of Fig.\ref{fig2} we report the behaviour of $P_>(k_{in})$ computed on the restricted nets. In this case the situation is very different, and no scale-free behaviour is observable. In particular, in US markets the maximum in-degree is significantly decreased, while in the Italian one it remains the same. This means that in the extended networks describing NYSE and NASDAQ the tail of $P_>(k_{in})$ is dominated by large investors outside the market, while in MIB it is dominated by listed companies, who are the largest holders of the market. For the small $k_{in}$ region of $P_>(k_{in})$ the opposite occurs. This is reflected in the fact that only $7\%$ of companies quoted in US markets invest in other companies, while the corresponding fraction is $57\%$ in the Italian case. 

\section{Pareto's law generalized to portfolio volume}
\begin{figure}[]	
\centerline{\epsfxsize 5.5 truein \epsfbox{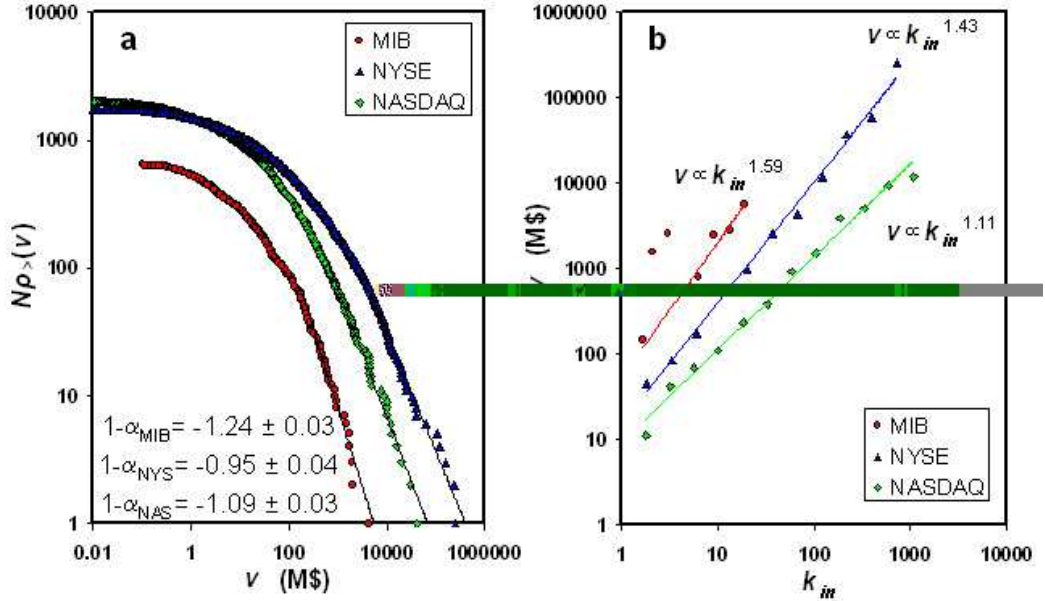}}
\caption[]{
\label{fig3}
\small a) Cumulative histrograms of $v$ (money units are millions of current US dollars, or $M\$$) for the extended nets and power-law fits to the tails. b) Scaling of $v$ against $k_{in}$. The straight lines are the curves $v(k_{in})\propto k_{in}^{1/\beta}$ with $\beta$ predicted by eq. (\ref{expon}), and are not fits to the data.}
\end{figure}

To capture the weighted nature of the networks, we also consider the number $\rho_>(v)$ of investors with portfolio volume greater than or equal to $v$. Once more (see Fig.\ref{fig3}a), we find that in all cases the tail of the distribution is well fitted by a power law 
\be\label{rho}
\rho_>(v)\propto v^{1-\alpha}
\ee
corresponding to a probability density
$\rho(v)\propto v^{-\alpha}$. The empirical values of the exponent are $\alpha_{NYS}=1.95$, $\alpha_{NAS}=2.09$, $\alpha_{MIB}=2.24$. Note that, since $v$ provides an estimate of the (invested) capital, the power-law behaviour can be directly related to the Pareto tails\cite{1,2,3,6} describing how wealth is distributed within the richest part of the economy. Consistently, note that also the small $v$ range of $\rho_>(v)$ seems to mimic the typical form displayed by the left part of many empirical wealth distributions\cite{3,6}, whose functional characterization is however controversial (log-normal, exponential and Gamma distributions have been equivalently proposed\cite{3,6} to reproduce it). Since in the following we are interested in the large $v$ and $k_{in}$ limit, the characterization of the left part of the distributions is however irrelevant, and we shall only consider the Pareto tails and the corresponding exponents. Note that, although the scale-free character encapsulated in eq. (\ref{pk}) is already known to be a widespread topological feature\cite{19}, power-law distributions describing the sum of link weights have only been addressed theoretically\cite{27} in the field of complex networks. Therefore our mapping of Pareto distributions (well established in the economic context\cite{1,2,3,6}) in a topological framework provides an empirical basis for the investigation of these specific properties of weighted networks.

\section{Scaling of portfolio volume versus portfolio diversification}
We now look for an additional characterization of our system. In particular, we ask if any relation between $(k_{in})_i$ and its weighted counterpart $v_i$ can be established.
If this is the case, then eqs. (\ref{pk}) and (\ref{rho}) are not independent since they can be derived from each other through the expression relating $v$ and $k^{in}$. In a topological context, this directly leads us to the framework explored in ref.\cite{fitness} where the degree of a vertex depends on an associated quantity or \emph{fitness}, which in this case is embodied in $v_i$. In such a case, the connection probability is necessarily fitness-dependent and its form -together with that of the fitness distribution- determines the topology of the network\cite{fitness}. 
Our empirical analysis reveals that this is indeed the case. As shown in Fig.\ref{fig3}b, we find that $v$ is an increasing function of the corresponding $k_{in}$, following an approximately straight line in double-logarithmic axes. The slope of this power-law curve is different across the three markets. However, in the Italian case two points deviate from this trend, signalling an anomalous behaviour for small ($k_{in}\le 3$) values of the diversification. We checked that these points correspond to investors holding a very large fraction ($\ge 50\%$) of the shares of an asset, whose portfolio has therefore a large volume even if its diversification is small. Clearly, these investors are the `effective controllers' of a company. While in both US markets the fraction of links in the network corresponding to such a large weight is of the order of $10^4$ (so that their effect is irrelevant on the plot of Fig.\ref{fig3}b), in MIB it equals the extraordinarily larger value $0.13$. This determines the `peak' at small $k_{in}$ superimposed to the power-law trend in the Italian market, and singles out another important difference between MIB and the US markets.

\section{The fitness model with generalized preferential attachment}
The results discussed so far are rather surprising since they show that portfolio structure is governed by simple laws in each of the three markets, allowing for an integrated description of both ordinary investors and companies despite their investments are expected to be driven by different factors. The former are in fact expected -at least within the standard framework of portfolio selection\cite{22}- to diversify their investments as much as possible in order to minimize financial risk, while companies instead organize their portfolios in a more focused way in order to establish strategic business alliances. 

Turning to a topological context, we now show that, as anticipated above, the observed properties can be reproduced by means of a recent stochastic network model\cite{fitness} that introduces a \emph{fitness} variable characterizing each vertex. Although the original model was designed for undirected graphs, it can be simply generalized to directed networks as follows. There are two types of vertices in the network, which in our case represent the $N$ agents (each characterized by its fitness $x_i$) and the $M$ assets (characterized by a different quantity $y_j$). Due to the presence of listed companies acting as both types, the total number of vertices does not sum up to $N+M$. We shall regard $x_i$ as proportional to the portfolio volume of $i$, which is the wealth that $i$ decides to invest. The quantity $y_j$ can instead be viewed as the information (such as the expected long-term dividends and profit streams) associated to the asset $j$. Note that $y_j$ can also be a vector of quantities, since the following results can be easily generalized to the multidimensional case\cite{pastor}. A link is drawn from $j$ to $i$ with a probability which is a function $f(x_i,y_j)$ of the associated properties. Note that $f(x,y)\ne f(y,x)$, differently from the undirected case\cite{fitness}. 

The simplest choice is the factorizable form $f(x,y)=g(x)h(y)$ where $g(x)$ is an increasing function of $x$, which takes into account the fact that investors with larger capital can afford larger information and transaction costs and are therefore more likely to diversify their portfolios. The function $h(y)$ encapsulates the strategy used by the investors to process the information $y$ relative to each asset. The stochastic nature of the model allows for two equally wealthy agents to make different choices (due for instance to different preferred investment sectors), even if assets with better expected long-term performance are statistically more likely to be chosen. For large web sizes, the expected in-degree of an investor with fitness $x$ is given by 
\be\label{kin}
k_{in}(x)=g(x)h_T
\ee
where $g_T$ is the total value of $g(x)$ summed over all $N$ agents. If the above relation is invertible, and if $\rho(x)$ denotes the statistical distributions of $x$ computed over the $N$ agents, then the in-degree distribution is given by \be\label{pkin}
P(k_{in})=\rho[x(k_{in})]dx(k_{in})/dk_{in}
\ee
Analogous relations for $k_{out}(y)$ and $P(k_{out})$ can be obtained directly. However, since our information regarding $k_{out}$ is incomplete (see above), we cannot test our model with respect to the function $h(y)$, and in the following we shall only consider the quantities derived from $g(x)$. 

Note that the above mechanism differs from those explored in most network models\cite{19}, where new vertices are continuously added and preferentially linked to pre-existing ones with large degree $k$ (`preferential attachment' rule). In the latter case, the functional form of the degree-dependent attachment probability can be measured\cite{19} in real evolving networks, and is found to be proportional to $k$ (`linear preferential attachment') or more generally to $k^\beta$ (`nonlinear preferential attachment'). Here, the attachment mechanism is `preferential' with respect to the variable $x$, and not to the pre-existing vertex degree. Within this `generalized preferential attachment' framework, the analogous choice for the connection probability is then $g(x)=cx^\beta$ with $\beta>0$, where $c$ is a normalization constant ensuring $0\le g(x)\le 1$ (a possible choice is $c=x_{max}^{-\beta}$, so that by defining $x\equiv v/v_{max}$ we can directly set $c=1$). It is straightforward to show that the predicted expressions (\ref{kin}) and (\ref{pkin}) now read  
\be\label{kin2}
k_{in}(x)\propto x^\beta
\ee
\be\label{pkin2}
P(k_{in})\propto k_{in}^{(1-\alpha-\beta)/\beta}
\ee
where we have used the fact that $\rho(x)\propto x^{-\alpha}$ for large $x$. Note that the above results still hold in the more general case when $f(x,y)$ is no longer factorizable provided that $k_{in}(x)=M\int f(x,y)\sigma(y)dy\propto x^\beta$ as in eq. (\ref{kin2}), where $\sigma(y)$ is the distribution of $y$ computed on the $M$ assets.

\section{Discussion and concluding remarks}
The empirical power-law forms of $\rho(x)$, $k_{in}(v)$ and $P(k_{in})$ are therefore in qualitative agreement with the model predictions. Moreover, by comparing eqs. (\ref{pk}) and (\ref{pkin2}) we find that the model predicts the following relation between the three exponents $\alpha$, $\beta$ and $\gamma$:
\be\label{expon}
\beta=(1-\alpha)/(1-\gamma)
\ee
By substituting in the above expression the empirical values of $\alpha$ and $\gamma$ obtained through the fit of Figs.\ref{fig2}a and \ref{fig3}a, we obtain the values of $\beta$ corresponding to the curves $v(k_{in})\propto k_{in}^{1/\beta}$ shown in Fig.\ref{fig3}b, which simply represent the inverse of eq.(\ref{kin2}) in terms of the quantity $v$. Remarkably, the curves are all in excellent agreement with the empirical points shown in the same figure, except the `anomalous' points of MIB. This suggests that the proposed mechanism fits well the investors' behaviour, apart from that of the effective holders of a company. 

A final comparison with the `traditional' preferential attachment mechanism is again revealing. Note that here we always observe the analogous of a superlinear ($\beta>1$) preferential attachment. However, while the traditional mechanism yields scale-free topologies only in the linear case\cite{19}, here we observe power-law degree distributions in the nonlinear case as well. This is a remarkable result, since in order to obtain the empirical forms of $P(k_{in})$ the exponent $\beta$ does not need to be fine-tuned, and the results are therefore more robust under modification of the model hypotheses. Also note that, interestingly, Pareto's law of wealth distribution has also been proposed\cite{2,30} as a possible explanation for the `fat tails' observed in financial markets fluctuations. In a network context, the above results support the hypothesis that the presence of non-topological quantities associated to the vertices may be at the basis of the emergence of complex scale-free topologies in a large number of real networks\cite{fitness}.

Authors acknowledge EU FET Open Project IST-2001-33555 COSIN for support and E. Sciubba, F. Lillo and M. Buchanan for helpful discussions and comments.

 
\end{document}